\documentclass[twocolumn,showpacs,preprintnumbers,amsmath,amssymb,APSl,prd,nofootinbib,superscriptaddress]{revtex4-2}

\usepackage{bm}
\usepackage{mathrsfs}
\usepackage{xcolor,color,graphicx,graphics}
\usepackage{epsfig,subfigure}
\usepackage{latexsym,amssymb,amsmath,amsfonts} 
\usepackage[english]{babel} 
\usepackage[OT1]{fontenc}
\usepackage[latin1]{inputenc}
\usepackage{makeidx}
\usepackage{hyperref}
\usepackage{color,graphicx,graphics,wrapfig,epsf}
\usepackage{hyperref}
\hypersetup{colorlinks=true, linkcolor=blue, citecolor=green}
\usepackage{makecell}
\usepackage{multirow}
\usepackage{booktabs}

\usepackage{appendix}
\usepackage{orcidlink}

\usepackage{lipsum}     
\usepackage{mathtools}

\usepackage{latexsym}
\usepackage{orcidlink}
\usepackage{enumerate}

\definecolor{red}{rgb}{1,0,0}

\def\+{^\dagger}

\def\<{\leftarrow}
\def\>{\rightarrow}

\def\({\left(}
\def\){\right)}

\newcommand{\bi}{\begin{itemize}} 				\newcommand{\ei}{\end{itemize}}
\newcommand{\benu}{\begin{enumerate}} 		\newcommand{\enu}{\end{enumerate}}
\newcommand{\bd}{\begin{dinglist}{0}}     \newcommand{\ed}{\end{dinglist}}
\newcommand{\bfig}{\begin{figure}[htbp]}  \newcommand{\efig}{\end{figure}}
        			
\newcommand{\bc}{\begin{center}} 				  \newcommand{\ec}{\end{center}}
\newcommand{\be}{\begin{equation}} 				\newcommand{\ee}{\end{equation}}
\newcommand{\bsub}{\begin{subequations}}  \newcommand{\esub}{\end{subequations}}
\newcommand{\ben}{\begin{eqnarray}} 			\newcommand{\een}{\end{eqnarray}}
\newcommand{\ba}[1]{\begin{array}{#1}} 		\newcommand{\ea}{\end{array}}
\newcommand{\bea}{\begin{equation}\begin{array}{rcl}}
\newcommand{\eea}{\end{array}\end{equation}}

\usepackage{soul}

\begin{document}
\title{The gravitational wave-black hole imaging correspondence for modified black holes}

\author{David D\'iaz-Guerra
        \orcidlink{0000-0003-0859-917X}
        }
        \email{ddiazgue@ucm.es}
        \affiliation{Departamento de F\'isica Te\'orica, Universidad Complutense de Madrid, E-28040 Madrid, Spain}
\author{\'Angel Rinc{\'o}n 
        \orcidlink{0000-0001-8069-9162}
        } 
        \email{angel.rincon@physics.slu.cz}
        \affiliation{Departamento de F{\'i}sica, Universidad del B{\'i}o-B{\'i}o,
Casilla 5-C, Concepci{\'o}n, Chile.}
        \affiliation{Research Centre for Theoretical Physics and Astrophysics, Institute of Physics, Silesian University in Opava, Bezrucovo n\'{a}m. 13, 74601 Opava, Czech Republic.}

\author{Diego Rubiera-Garcia 
        \orcidlink{0000-0003-3984-9864}
        }
        \email{drubiera@ucm.es}
        \affiliation{Departamento de F\'isica Te\'orica and IPARCOS, Universidad Complutense de Madrid, E-28040 Madrid, Spain}

        \author{Diego Saez-Chillon Gomez
        \orcidlink{?}
        }
        \email{diego.saez@uva.es}
        \affiliation{Department of Theoretical, Atomic and Optical Physics,
and Laboratory for Disruptive Interdisciplinary Science (LaDIS), Campus Miguel Delibes, \\ 
University of Valladolid UVA, Paseo Bel\'en, 7, 47011 Valladolid, Spain}
\affiliation{Department of Physics, Universidade Federal do Cear\'a (UFC), Campus do Pici, Fortaleza - CE, C.P. 6030, 60455-760 - Brazil}

\date{\today}
\begin{abstract}

Black holes (BHs) can be studied via fundamentally different observational channels that probe complementary aspects of their physics. While BH imaging provides access to the quasi-static space-time geometry via the strong bending of light rays, gravitational wave (GW) observations probe the dynamical response of the space-time to time-dependent processes in the inspiral, merger and ringdown phases. Both messengers -- electromagnetic imaging probes and ringdown GW spectroscopy --, provide access to essentially the same region -- the one between the BH event horizon and the photon region --, but they do it via conceptually different methods, encoding different physical information. However, it has been shown in the literature that physical quantities supposedly exclusive of each such messenger are actually tightly related to each another via a correspondence that occurs in the eikonal limit (i.e. large values of the multipole number $\ell$) of the geometric-optics approximation.  In this paper we clarify the actual identification of observables within such a correspondence and test its accuracy for  modified spherically symmetric BH geometries proposed in the literature. We find that even for low values of $\ell$ the correspondence is surprisingly accurate in relating the real and imaginary parts of quasi-normal modes in the GW ringdown phase with the critical impact parameter and Lyapunov exponent of nearly-bound light trajectories for every such model analyzed. We discuss the applicability of such a result both for each messenger individually, and also for foreseeable tests of BHs combining both messengers.

\end{abstract}

\maketitle

\section{Introduction}

The existence of black holes (BHs) is one of the pivotal predictions of our theories of the gravitational field. Within Einstein's General Relativity (GR), such BHs are described by the Kerr solution \cite{Kerr:1963ud}: the most general asymptotically-flat, stationary, axi-symmetric, vacuum solution of Einstein's field equations, solely described by mass and angular momentum \cite{Robinson:1975bv}. However, BHs departing from the Kerr solution can be found via the introduction of additional ``charges", such as electromagnetic \cite{Newman:1965my} or scalar \cite{Herdeiro:2014goa} fields, or by enlarging the framework of GR within modified theories of gravity \cite{Sotiriou:2008rp,Olmo:2011uz,Clifton:2011jh,Nojiri:2017ncd}. Some of such modified BHs are capable of displaying theoretically consistent physical features while at the same time can be studied using the same tools employed for the Kerr metric \cite{Berti:2015itd,Yunes:2013dva}.

Currently, the two main messengers at our disposal for testing whether the nature of BHs is accurately described by the Kerr solution or instead traces of non-Kerrness are present, are sensitive to the same BH parameters but they do it via observational routes that differ substantially. On the one hand, BH imaging (BHI) tests the background space-time geometry under quasi-stationary conditions, focusing on the images brought by highly-bent trajectories near the photon region \cite{Cunha:2018acu}. On the other hand, gravitational waves (GWs) probe the temporal evolution of the space-time, appealing to the highly non-linear dynamics of the horizon formation and relaxation into the BH final stationary state, reaching their appex in BH ringdown spectroscopy \cite{Berti:2009kk}. From this perspective, one could say that while BHI provides a snapshot of the BH geometry, GWs supply a movie of the space-time in evolution. 

With the detection of GWs out of binary BH mergers by the LIGO-Virgo Collaboration \cite{LIGOScientific:2016aoc,LIGOScientific:2017vwq}, and with the successful imaging of the light emitted by the hot material orbiting M87$^*$ \cite{EventHorizonTelescope:2019dse} and Sgr A$^*$ \cite{EventHorizonTelescope:2022wkp}
by the Event Horizon Telescope (EHT), we have opened a window allowing us to probe  the strong-field regime of BHs via the above two messengers.  Further planned observatories, including the next-generation EHT \cite{Tiede:2022grp}, the BH Explorer \cite{Johnson:2024ttr}, LISA \cite{Barausse:2020rsu}, or the Einstein Telescope \cite{ET:2025xjr}, engender further promises for high-precision BH physics.

Within this context, ideas about combining information from BHI and GWs probes for investigating both stationary and dynamical aspects of BHs using tools from each messenger are currently gaining traction. One such idea was suggested decades ago \cite{Ferrari:1984zz,Cardoso:2008bp} but it has recently re-emerged with renewed vigor, and for the sake of this paper we shall dub it as the quasi-normal modes-black hole imaging (QNM-BHI) correspondence. It is grounded on the fact that not only do BHI and GWs (in the ringdown phase) probe essentially the same space-time region  -- the one comprised between the BH event horizon and the photon (shell) region -- but they do it using a description in terms of effective potentials that closely resemble one another. In fact, in the geometrical-optics approximation given by the eikonal limit -- that is, large values of the multipole number $\ell$ -- such a correspondence is exact. On the one end of such a correspondence are the real and imaginary parts of the QNMs of GWs found in the relaxation (ringdown) phase after the coalescence has taken place, and which are physically interpreted as the frequency and damping rate of the QNMs, respectively. On the other end we find the critical impact parameter and the Lyapunov index of nearly-bound geodesics of highly-bent light rays, whose physical meaning can be linked with the central dark region and bright photon ring features appearing in the image, respectively. While the QNM-BHI correspondence is formally valid in the eikonal regime only \cite{Cardoso:2008bp}, there has been some sparse works in the literature suggesting that, for particular metrics, it might also provide good numerical estimates for quantities on each side for low $\ell$ \cite{Konoplya:2019xmn,Yu:2022yyv,Pedrotti:2025idg}.

The main aim of the present paper is to test the actual usefulness of the correspondence to provide the correct numbers of such quantities from one of its ends to the other (and the other way round) for modified BH configurations. The latter correspond to a set of ten static spherically symmetric configurations, selected from the pool of sixteen geometries considered in Ref. \cite{daSilva:2023jxa} (and which are actually a subset of the $\sim 60$ geometries considered in the highly influential work \cite{Vagnozzi:2022moj}) on the grounds of easiness of the numerical procedures on both the GW and BHI sides. While such geometries can be supported by modifications of the Schwarzschild solution of GR either via the introduction of additional matter fields or framed with modified gravity theories, we shall remain agnostic about their origin, but instead focus on the way the correspondence is realized for different values of $\ell$. Surprisingly, we find that the correspondence works remarkably well even for low values of $\ell$. This means that starting from (say) the computation of QNM frequencies we get the quantities characterizing the BHI side (critical impact parameter and Lyapunov exponent) with little loss of accuracy as compared to their values found through usual ray-tracing procedures of the BHI field. 

We shall elaborate on the caveats regarding the connection between these theoretical quantities against the actual modeling and observational difficulties in measuring them for each messenger. It should be pointed out the existence of previous works on the relation between QNMs and BHI quantities for different specific models, see e.g. \cite{Konoplya:2017wot,Jusufi:2020odz,Liu:2020ola,Li:2021zct,Okyay:2021nnh,Campos:2021sff,Guo:2021bcw,Pantig:2022gih, Atamurotov:2022nim, Pedrotti:2024znu,Lambiase:2024lvo,Bonanno:2025dry,Borah:2025tvw,Feng:2025iao}.

This paper is organized as follows. In Sec. \ref{S:II} we discuss the physical origin and interpretation of QNMs as well as the numerical methods employed to find them. In Sec. \ref{S:III} we consider the process of  generation of BH images and the quantities associated to nearly-bound geodesics. The QNM-BHI correspondence is introduced in Sec. \ref{S:IV}, where we supply a physical interpretation to quantities on each side and across the divide between them, clarifying some misunderstandings in the literature of the topic. In Sec. \ref{S:V} we introduce a bunch of non-Schwarzschild geometries, and compute their BHI observables and QNM frequencies for picks on the metric coefficients that fall within some recent constraints from the EHT Collaboration on M87$^*$ shadow's size. We discuss the degree to which the correspondence allows to supply the right numbers for the critical impact parameter and the Lyapunov exponent of nearly-bound geodesics from the real and imaginary parts of the QNM frequencies at different values of the multipole number $\ell$. Furthermore we discuss the usefulness of the correspondence when framed within the theoretical and observational difficulties associated to the measurements of the physical features ascribed to each messenger.  We conclude in Sec. \ref{S:VI} with some further discussion.

\section{Quasi-normal modes} \label{S:II}

QNMs of BHs are the characteristic damped oscillations that dominate the ringdown phase of perturbed BH space-times. They describe how a BH returns to equilibrium after a dynamical disturbance, such as a binary merger, accretion event, or external perturbation. In a simplified picture, a perturbed BH ``rings" like a damped bell, emitting gravitational waves (or other fields) at specific complex frequencies
\begin{align}
\omega &= \omega_R + i \omega_I \quad (\omega_I < 0),
\end{align}
where: 
(i) the real part $\omega_R$ determines the oscillation frequency, and, 
(ii) the negative imaginary part $\omega_I$ governs the exponential decay rate.

QNMs (of BHs) arise from the application of linear perturbation theory in GR. For the sake of this paper we consider a spherically symmetric space-time whose line element, in four space-time dimensions, is of the form
\begin{equation} \label{eq:lineel}
ds^2=-A(r)dt^2 + B(r)dr^2 + r^2(d\theta^2 + \sin^2 \theta d\varphi^2),
\end{equation}
where $\{A(r),B(r)\}$ are the metric functions. Typically, after the separation of the wave function into its temporal, angular (spherical harmonics) and radial parts, a Schr\"odinger-like equation for the latter of the following form is written:
\begin{equation} \label{eq:weq}
\psi_{yy} + \left( \omega^2 - U(y) \right)\psi=0
\end{equation}
where $\omega$ denotes the wave's frequency, and $U\equiv U(y)$ is the effective potential written in terms of  some radial variable $y$, generally a tortoise-like coordinate, with $\psi_{y}\equiv d\psi/dy$.  Such variable maps the external region of the BH horizon $r_h$ in the radial coordinate $r \in (r_h,\infty)$ into $y \to (-\infty,+\infty)$.  To mathematically close the problem one needs to set suitable boundary conditions, which could vary slightly, for example, by including a massive scalar field. These correspond to the requirements of purely ingoing modes at the event horizon and purely outgoing at asymptotic infinity, namely
\begin{align}
\psi(y) &\sim e^{-i \, \omega \, y}, \quad y \to -\infty \\
\psi(y) &\sim e^{+i \, \omega \, y}, \quad y \to +\infty 
\end{align}
interpreted as the lack of reflected waves at the BH event horizon and the lack of waves coming from external sources, respectively. For instance, in the case of scalar perturbations and when the Schwarzschild gauge is assumed (i.e., $g_{rr}g_{tt}=-1$), the corresponding potential for many spherically symmetric black holes can be cast under the form
\begin{equation} \label{eq:UQNM}
U_{QNM}= A(r) \left[\frac{\ell (\ell+1)}{r^2} + \frac{A'(r)}{r} \right],
\end{equation}
where $\ell$ is the multipole number.

The foundational application of this framework was the Schwarzschild BHs, where scalar, electromagnetic, and gravitational perturbations are governed by the well-known Regge-Wheeler equation (for odd-parity modes) and Zerilli equation (for even-parity modes) \cite{Regge:1957td,Zerilli:1970se}. In the case of rotating Kerr BHs, the Teukolsky equation provides the appropriate formalism for spin-weighted perturbations \cite{Teukolsky:1973ha}. While this classical picture remains the standard paradigm, recent works have raised subtle questions regarding the precise definition and interpretation of QNMs suggesting that  precision in terminology may be required \cite{Steinhauer:2025sqd}.

QNMs can be obtained in a plethora of ways, including not only exact, but also analytic and semianalytic methods. The most common of these are the WKB approximation (particularly powerful in the eikonal limit $  \ell \gg 1  $) \cite{Konoplya:2011qq,Berti:2009kk}, continued-fraction techniques \cite{Leaver:1985ax}, and pseudospectral methods \cite{Jansen:2017oag}. These approaches allow high-precision determination of the complex spectrum and have been extended to a wide range of BH geometries and modified gravity theories.

Determining the QNM of a given process is of the utmost relevance since they play a central role in a broad set of applications in modern astrophysics and fundamental physics. For example, they allow for BH spectroscopy from gravitational-wave ringdown signals, enabling precise measurements of mass, spin and possibly charge \cite{Berti:2009kk,Konoplya:2011qq}. They are also key probes of modified gravity and exotic compact objects \cite{Konoplya:2011qq}, and have been at the center of tests of the no-hair theorem in the strong-field regime \cite{Isi:2019aib}.

\section{BH imaging} \label{S:III}

The Lagrangian that describes the motion of a photon along an arbitrary geodesic trajectory $x^{\mu}(s)$, where $s$ is the affine parameter, is defined as
\begin{equation} \label{eq:geo}
\mathcal{L}=g_{\mu\nu}\dot{x}^{\mu}\dot{x}^{\nu}\equiv 0,
\end{equation}
where a dot denotes a derivative with respect to the affine parameter, i.e., $\dot{x}^{\mu} \equiv dx^{\mu}/ds$. This implies that photons follow null geodesics on the background geometry\footnote{Note that, if the geometry is supported by models of non-linear electrodynamics (NEDs), then photons travel through geodesics of an effective geometry instead \cite{Novello:1999pg,DeLorenci:2000yh}, while cosmological effects can also affect photon propagation \cite{Odintsov:2022umu}.}.  The Euler-Lagrangian equations allows to obtain the associated momenta of the system, $p_{\mu} = \partial \mathcal{L}/\partial \dot{x}^{\mu} = \dot{x}_{\mu}$, and the static and spherically symmetric nature of the line element means that there are two such momenta which are conserved: the energy (per unit mass)
\begin{equation}
p_t=-A \dot{t} \equiv E, \label{eq:energy}
\end{equation}
and the azimuthal angular momentum
\begin{equation} \label{eq:momentum}
p_{\varphi}=r^2 \sin^2 \theta \dot{\varphi}.
\end{equation}
Note that due to the spherical symmetry of the system, we can pick the plane $\theta=\pi/2$ for casting the corresponding equations without any loss of generality, which greatly simplifies the analysis.

Under these conditions, the geodesic equation (\ref{eq:geo}) can be written as 
\begin{equation} \label{eq:BHIscat}
AB \dot{r}^2=\frac{1}{b^2}-V(r),
\end{equation}
where a factor $L^2$ has been re-absorbed in the definition of the affine parameter, $s \to s/L^2$, and we have introduced the definition
\begin{equation}
V(r)=\frac{A(r)}{r^2},
\end{equation}
which is the effective potential for BHI (observe the resemblance to the potential of the QNM problem, Eq.(\ref{eq:UQNM})).  Here $b \equiv L/E$ is the normalized angular momentum and can be identified as the apparent photon's {\it impact parameter} on the observer's screen.

It is possible for photons to move in circular geodesics, $\dot{r}=0=\ddot{r}$, for which the effective potential provides the expressions $\{V(r)= 1/b^2; V'(r)=0\}$. The second condition is conveniently written as
\begin{equation} \label{eq:cir1}
    \left[ \frac{A'}{A} - \frac{2}{r} \right]_{r = r_{ps}}=0 \rightarrow r_{ps}=\frac{2A_{ps}}{A'_{ps}}.
\end{equation}
The solution (if any) $r=r_{ps}$ is dubbed as the {\it photon sphere}. This way, the first condition above provides the {\it critical impact parameter} associated to the photon sphere as
\begin{equation} \label{eq:cir2}
b_{ps}= \frac{r_{ps}}{\sqrt{A_{ps}}} \rightarrow b_{ps}=\frac{2A_{ps}^{1/2}}{A_{ps}'},
\end{equation}
where the subscript ``PS" means evaluation at the photon sphere, e.g. $A_{ps} \equiv A(r_{ps})$.  It is worth pointing out that BHs always have at least a photon sphere \cite{Carballo-Rubio:2024uas}.  

The photon sphere is the locus of unstable bound geodesics, while the critical curve corresponds to its projection on the observer's screen, taking the shape of a perfect concentric circle\footnote{For rotating BHs the photon sphere is enlarged into a {\it photon shell} defined by the set of bound librations between a pair of radial $r \in (r_-,r_+)$ and angular $\theta \in (\theta_-, \theta_+)$ regions.}. To understand its unstable character, let us consider a photon that starts its trip at some location very close to the photon sphere, that is, $r=r_{ps}+\delta r_0$. Expanding the effective potential in (\ref{eq:BHIscat}) we get
\begin{align}
\begin{split}
V(r) \approx & \ V(r_{ps}) + V'(r)\vert_{r=r_{ps}} \delta r \ 
\\
+&  \ \frac{1}{2}V''\vert_{r=r_{ps}} \delta r^2 + \mathcal{O}(\delta r^3),
\end{split}
\end{align}
where the term $V'(r)\vert_{r=r_{ps}}$ vanishes due to the photon sphere condition, while we have
\begin{equation} \label{eq:inter}
V''\vert_{r=r_{ps}}=\frac{1}{r_{ps}^4} \left( r_{ps}^2 A_{ps}''-2A_{ps} \right).
\end{equation}
Replacing the above expansion into the geodesic equation (\ref{eq:BHIscat}), using the angular momentum conservation equation, and removing zeroth-order terms we get, at second-order
\begin{equation}
\frac{A_{ps}B_{ps}}{r_{ps}^4} \left(\frac{d\delta r}{d \delta \varphi}\right)^2=-\frac{1}{2} V''\vert_{r=r_{ps}} \delta r^2,
\end{equation}
which can be suitably re-written as
\begin{equation} \label{eq:nerboun1}
\pi \left(\frac{d \delta r}{d\varphi}\right)= \gamma_{ps} \delta r ,
\end{equation}
which can be integrated as 
\begin{equation}
\delta r=\delta r_0 e^{\gamma_{ps} \tfrac{\varphi}{\pi}} ,
\end{equation}
so that by introducing nearly-bound quantities as
\begin{equation}
\tilde{r}=\frac{r}{r_{ps}}-1 \quad ; \quad
\tilde{b}= \frac{b}{b_{ps}}-1,
\end{equation}
then the location of the photon with a given impact parameter $\tilde{b} \ll 1$, after an angle $\varphi$ has elapsed, reads as
\begin{equation} \label{eq:lyaexp}
\tilde{r}=\tilde{r}_0 e^{\gamma_{ps} \tfrac{\varphi}{\pi}} ,
\end{equation}
where $\tilde{r}_0=\tilde{r}(\varphi=0)$ is initial photon's location. In the above expression, the quantity 
\begin{equation} \label{eq:lyaexpp}
\gamma_{ps}=\pi \frac{1}{A'_{ps}B_{ps}^{1/2}} \left(A_{ps}^{'2}-2A_{ps} A_{ps}'' \right)^{1/2} 
\end{equation}
is dubbed as the {\it lensing Lyapunov exponent}. This exponent controls the exponential drift of nearly-bound geodesics in the azimuthal angle. Furthermore, by introducing the number of half-turns 
\begin{equation}
n \equiv \frac{\varphi}{\pi}
\end{equation}
the set of geodesic trajectories can be classified into {\it lensing bands}, that is, light rays that have completed at least $n$ half-turns around the BH (i.e. at least $(n+1)$ crossings with the equatorial plane $\theta=\pi/2)$ before escaping to asymptotic infinity or falling into the BH event horizon~\cite{Gralla:2020srx}. This way, light rays with $n=0$ correspond to the disk's direct emission, while those with $n=1,2, \ldots$ lie entirely within the $n^{th}$ lensing band and will produce, on the observer's screen, a sequence of {\it photon rings} indexed by $n$, and surrounding a central dark region.

It is possible to rewrite the nearly-bound orbit equation in terms of the coordinate time. To this end, we simply use the conservation of the angular momentum equation (\ref{eq:momentum}) into Eq.(\ref{eq:nerboun1}) and, after some suitable manipulations, we arrive at
\begin{equation} \label{eq:lyatime}
\tilde{r}=\tilde{r}_0 e^{\lambda_{ps} t} ,
\end{equation}
where $\tilde{r}_0 \equiv \tilde{r}(t=0) \ll 1$ and 
\begin{equation} \label{eq:lyaprin}
\lambda_{ps}=\frac{\gamma_{ps}}{\pi b_{ps}} ,
\end{equation}
is dubbed as the {\it principal Lyapunov index}. This equations tells us about the drift of radial distance after a coordinate time $t$ has elapsed. Therefore, it defines a characteristic time, $t_{ps}=\lambda_{ps}^{-1}$, which has physical relevance in transient observations such as hot-spots. Note that this exponent was derived already by Cardoso in \cite{Cardoso:2008bp} using a complementary method.

For further reference, for a Schwarzschild space-time these relevant quantities for BHI read as 
\begin{equation}
r_{ps}=3M; \quad b_{ps}=3\sqrt{3}M; \quad 
\gamma_{ps}= \pi; \quad \lambda_{ps}=\frac{1}{3\sqrt{3}M}.
\end{equation}

For the sake of our work we shall give attention to $\gamma_{ps}$ (rather than $\lambda_{ps}$), since it is the one that can be directly compared with the observables in time-averaged images of a BH.

\section{QNM-BHI Correspondence} \label{S:IV}

In this section we provide first an overview of the origin and later the physical implications of the QNM-BHI correspondence. 

\subsection{The correspondence in the geometrical-optics approximation}

The correspondence has its basis on the geometrical-optics approximation, in which the wavelength $\lambda$ of the test field (scalar, electromagnetic, etc) is much shorter than any other length-scale $L$ present in the system, in other words 
\begin{equation}
\nabla^{\mu} \nabla_{\mu}\textbf{A}=\mathcal{O}(\lambda/L) \sim 0,
\end{equation}
for a given test field $\bf{A}$, and therefore in this {\it eikonal limit} this will be the single operator that dominates the dynamics of the system over any other field/interaction. In this limit, the wave solution of the above equation naturally delivers (at leading order) the equation of light rays propagating in the background space-time, $k^{\mu}\nabla_{\nu}k_{\mu}=0$, where $k_\mu= \partial_\mu S$ is both the photon's wave number and the test field phase $S$.

The QNM-BHI correspondence therefore stems from considering wave propagation of a test-field in a spherically symmetric space-time following the GW equation (\ref{eq:weq}).  Let us assume the test field to take the usual ansatz given by (for a more detailed derivation see e.g. \cite{Glampedakis:2019dqh,Chen:2022ynz})
\begin{equation}
\psi(y)=a(\vert y \vert) e^{iS(y)/\epsilon},
\end{equation}
for the modulus $a$ and the phase $S$. Furthermore, $S$ varies on the length scale of the field wavelength, while the corresponding amplitude changes {\it{slowly}}.
Here $\epsilon$ tracks the order in the eikonal approximation. In the leading-order of such an approximation, the wave equation (\ref{eq:weq}) implies that $\omega^2$ must be real, and therefore $\omega$ can only be real or imaginary. Assuming the former, then the frequency is expanded as
\begin{equation} \label{eq:freex}
\omega= \omega_R^{(0)} + \omega_R^{(1)} + i \omega_I^{(1)} + \mathcal{O}(2),
\end{equation}
where $R/I$ stand for real/imaginary parts, and the subindices $(0)$ and $(1)$ stand for the order in the approximation. This way, the leading-order frequency term of the wave equation can be solved as
\begin{equation} \label{eq:realpart}
\omega_R^{(0)}= \sqrt{U_m},
\end{equation}
where ``$m$" denotes the evaluation of the effective potential at the maximum of the potential $y_m\equiv y(r_m)$, i.e., $U_m \equiv U(y_m)$. At the next-to-leading order, it is possible to recover the expression of the first contribution to the imaginary part as \cite{Chen:2022ynz}
\begin{equation}
\omega_I^{(1)}= -\frac{1}{2} \sqrt{\frac{-U_{yy,m}}{2U_m}}.
\end{equation}
Now let us assume that our wave equation is such that it can be written under the form
\begin{equation} \label{eqPoteik}
U(y(r))=\frac{\ell(\ell+1)}{r^2}A(r) + \mathcal{O}(\ell^0),
\end{equation}
which is the typical structure found in the Schwarzschild solution and in many other known modified BH solutions in which the Schwarzschild gauge is assumed. Given the fact that in the eikonal limit the first term dominates over the other ones, we can just ignore the latter. From Eq.(\ref{eq:realpart}), one therefore finds that, in this limit:
\begin{equation} \label{eq:impart}
\omega_R^{(0)}= \left(\ell+\frac{1}{2} \right) \frac{A_m^{1/2}}{r_m},
\end{equation}
and from (\ref{eq:impart}) that
\begin{equation} \label{eq_omega1}
\omega_I^{(1)}=-\frac{1}{2}\sqrt{- \frac{r^2}{A}U_{yy}} \Big \vert_{r=r_m}.
\end{equation}
To complete the QNM-BHI correspondence we now call upon the WKB approximation, since in the eikonal limit it provides semi-analytic expressions for the complex QNM frequencies. The QNM condition of the WKB method provides the relation
\begin{equation}
\frac{Q_m}{\sqrt{2Q_{yy}(y_m)}}=i \left(n+\frac{1}{2} \right),
\end{equation}
where $Q=\omega^2-U(y)$. In the eikonal limit, where the effective potential is given by (\ref{eqPoteik}), this equation is cast as
\begin{equation}
\omega^2=\ell(\ell+1)\frac{A_m}{r^2_m} + i \left(n+\frac{1}{2} \right) \sqrt{-2\ell(\ell+1) \frac{d^2}{dy^2} \left(\frac{A}{r^2} \right)} \Big \vert_{r=r_m}.
\end{equation}
Using now Eq.(\ref{eq:freex}) one finds
\begin{equation}
\omega_{QNM}=\omega_R +i \omega_I=\left(\ell+\frac{1}{2} \right) \Omega_R - i\left(n+\frac{1}{2}\right) \Omega_I,
\end{equation}
where
\begin{equation}
\Omega_R=\frac{A_m^{1/2}}{r_m} \quad ; \quad \Omega_I= \frac{1}{\sqrt{2}} \sqrt{-\frac{r_m^2}{A} U_{yy}},
\end{equation}
which closes our summary on the origin of the correspondence.

\subsection{Interpretation of quantities}

We can proceed now with the interpretation of the above quantities within each side of the correspondence.

\subsubsection{QNM real part vs shadow}

For the real part of the QNM frequency, $\omega_R=(\ell+\tfrac{1}{2})\Omega_R$, one first notes that the radius $r_m$ is identified with the photon sphere radius, $r_m=r_{ps}$, as follows from the fact that in this eikonal limit, the location of the maximum of the QNM potential is the same as in the effective potential of BHI. On the other hand, one can verify that $\Omega_R \equiv \tfrac{\dot{\phi}}{\dot{t}}$, that is, it is simply the angular velocity of unstable bound geodesic, and furthermore it coincides with the (inverse) critical impact parameter (in the radial coordinate system of Eq.(\ref{eq:cir2})). Explicitly 
\begin{equation} \label{eq:correspondence1}
\omega_R=\frac{\left(\ell+\tfrac{1}{2}\right)}{b_{ps}}.
\end{equation}
From a physical perspective, the real part $\omega_R$ thus corresponds to the orbital frequency of the unstable circular null geodesic at the photon sphere. Given this connection it is tempting to associate such a QNM wave amplitude (i.e. $\omega_R$) to the (inverse) shadow's radius, since in the usual shadow interpretation, as provided by Falcke's view \cite{Falcke:1999pj}, the dark central region of BHI is bounded by light rays that asymptote to the critical curve. However, current knowledge on BHI has revealed that in non-spherical accretion disks, i.e., as long as there are gaps in the emission region, and for an accretion flow that extends up to the event horizon, which is the actual physical scenario there will be contributions to the image brightness appearing inside the critical curve \cite{Chael:2021rjo,Vincent:2022fwj,Cardenas-Avendano:2023dzo}. This breaks the simple association between shadow (i.e. the critical curve) and the actual brightness deficit and, hence, with the QNM oscillation frequency.

\subsubsection{QNM imaginary part vs photon rings}

To interpret $\Omega_I$ we must sweat a little more. We first note that the coordinate $y$ of the QNM problem should be related to the radial one $r$ of the BHI one, so we can write 
\begin{equation}
U_{yy}=U_{rr} \left( \frac{dr}{dy} \right)^2.
\end{equation}
Typically, the coordinate in which the wave problem is formulated corresponds to the tortoise coordinate, which in the line element (\ref{eq:lineel}) takes the form (for trivial radial function)
\begin{equation} \label{eq:tortoise}
\frac{dr}{dy}=\sqrt{\frac{A}{B}},
\end{equation}
which implies that
\begin{equation}
U_{yy}\Big\vert_{ps}=\frac{A_{m}}{B_{m}}U_{rr} \Big \vert_{ps},
\end{equation}
so that we have
\begin{equation}
\Omega_I= \frac{1}{\sqrt{2}} \sqrt{-\frac{r_{ps}^2}{B} U_{rr}} .
\end{equation}
We need another intermediate relation, given by (which is simply Eq.(\ref{eq:inter}) replacing $V \to U$)
\begin{equation}
  U_{rr} =  \frac{d^2}{dr^2} \left(\frac{A}{r^2} \right)=\frac{1}{r^4} \left(r^2 A_{rr}-2A \right),
\end{equation}
so that
\begin{equation}
\Omega_I=\frac{1}{\sqrt{2}} \sqrt{-\frac{1}{Br^2} (r^2 A_{rr}-2A )} \Big \vert_{ps}.
\end{equation}
Now, applying the relations (\ref{eq:cir1}) and (\ref{eq:cir2}), and comparing with Eqs.(\ref{eq:lyaprin}) and (\ref{eq:lyaexpp}) one concludes that
\begin{equation}
\Omega_I=\lambda_{ps},
\end{equation}
i.e., this is just the principal Lyapunov index. Therefore one has
\begin{equation} \label{eq:correspondence2}
\omega_I= \left(n+\frac{1}{2}\right) \vert \lambda_{ps} \vert  .
\end{equation}
This equation links the imaginary part of the QNM, $\omega_I$, with the Lyapunov exponent characterizing the geodesic instability \citep{Konoplya:2011qq,Yu:2022yyv,Cardoso:2008bp}. It is therefore tempting to associate the damping time of the QNM with the time-scale of unstable bound orbits. From the perspective of BH imaging within EHT observations, the current best tool at of our disposal is the measurement of time-averaged images, since this is where the photon rings manifest. This simply means using the lensing Lyapunov exponent (\ref{eq:lyaexpp}) via the relation (\ref{eq:lyatime}) to write
\begin{equation}
\gamma_{ps}=\frac{\pi b_{ps} \omega_I}{n+\frac{1}{2} }
\end{equation}
which would link the QNM damping rate/time with physical properties associated to photon rings.  It is important to stress, however, that the Lyapunov index implicitly assumes a completely homogeneous medium seen by photons as they travel throughout the disk. However, this is far from the reality of accretion disks, in which the turbulent, highly-magnetized environment entails significant differences in the regions traveled by photons labeled by a number of half-orbits $n$, rendering this connection between $\omega_I$ and photon rings more subtle, as shall be discussed later on.

\subsection{The role of the WKB method in the correspondence}

Let us further emphasize, for the sake of completeness, the now evident connection not only between QNMs and BHI, but also their intimate relationship with the well known WKB approximation. The cornerstone of this connection lies in the relationship between geodesic optics and wave scattering, in the eikonal regime, characterized by $ \ell \gg 1 $ (and typically $ \ell \gg n $, where $ n $ is the overtone number). This interplay reveals how null geodesics (particularly the unstable photon sphere) govern both the characteristic ringing frequencies and the critical quantities that determine the BH features in the image. This correspondence arises because the effective potential barrier in the radial perturbation equation reaches its maximum near the photon sphere, causing high-frequency waves to behave as localized wave packets propagating along null geodesics.

Mashhoon's pioneering work on the analytic description of BH oscillations demonstrated long ago that, in the eikonal limit, QNM are intimately connected to the properties of unstable photon orbits \cite{Mashhoon:1985cya}. A more complete and detailed analysis of the geodesic correspondence was later presented by Cardoso \textit{et al.} \cite{Cardoso:2008bp}. More recently, several important results have emerged in the eikonal regime, including the relation between the real part of the QNM and the radius of the BH shadow \cite{Cuadros-Melgar:2020kqn}. 

The WKB approximation enters naturally here. Schutz and Will \cite{Schutz:1985zz} first applied it to BH QNMs, deriving a semi-analytic formula that becomes exact in the eikonal limit. The higher-order extensions by Iyer and Will \cite{Iyer:1986np}, later Konoplya up to sixth order \cite{Konoplya:2003ii} and Matyjasek and Opala up to order 13 \cite{Matyjasek:2017psv}, provided practical tools for computing QNMs. In this framework, the QNM frequency satisfies a quantization condition derived from matching asymptotic solutions across the potential barrier. In the large-$\ell$ limit, the dominant contribution yields the direct relation $ \omega_R \approx \ell \Omega_R $, while the imaginary part is governed by the curvature of the effective potential at its maximum, namely the second derivative at the peak, which is associated with the orbital instability timescale $ |\lambda_{ps}^{-1}| $. This WKB-derived structure underpins the explicit QNM-BHI correspondence. In modified gravity theories and higher-dimensional scenarios, the same general framework remains valid. For example, in string-inspired or Einstein-Gauss-Bonnet BHs \cite{Paul:2023eep}, both the properties of the photon sphere -- and consequently the features characterizing BHI -- and the eikonal QNMs depend sensitively and coherently on the underlying coupling parameters, with the WKB approach providing reliable and consistent approximations.

The physical picture behind this correspondence is rather natural, since both effects are governed by the same unstable photon orbit. On one hand, the BHI arises from the behavior of null geodesics in the geometric-optics limit. On the other hand, eikonal QNMs describe the oscillatory response of perturbations localized near that unstable orbit. The WKB approximation connects these two descriptions by providing a semiclassical treatment valid precisely in the high-frequency regime where geometric optics becomes applicable. Of course, this correspondence could fail and therefore it is not guaranteed to be valid a priori. Thus, the approximation should become less accurate for small values of $ \ell $ (or large overtone numbers $ n $), while strong deviations from spherical symmetry may lead to mode mixing. To analyze its usefulness, in the next section we check its degree of accuracy to relate each quantity on each side of the correspondence for a bunch of modified spherically symmetric BH geometries at both large and small multipole numbers $\ell$.

\section{The validity of the correspondence for modified BH geometries} \label{S:V}

In order to break the assumptions of the uniqueness theorems on BHs within GR, which single out the Kerr metric, two paths are typically followed: either the introduction of matter fields (electromagnetic, scalar), or extending the gravitational framework (modified theories of gravity). Either way, the resulting modified BH will be characterized by an additional set of parameters encoding the contributions of either the additional matter fields or the new  gravitational interactions. For the sake of this work, we shall consider a bunch of well known solutions in the literature characterized by a single or two parameters.

\subsection{Modified BH models}

Our selection of models is grounded on the analysis of \cite{daSilva:2023jxa}, where the authors study different modified BH geometries inherited from the far larger selection considered in \cite{Vagnozzi:2022moj}. The latter reference describes how the constraints found by the EHT Collaboration in \cite{EventHorizonTelescope:2022xqj} on the shadow's size (in Falcke's sense \cite{Falcke:1999pj}) are translated into constraints for the parameters of spherically symmetric BH metrics, while the former reference computes the Lyapunov exponent for sixteen of them. For the purposes of the present work, we pick ten of them, selected according to their easiness in implementing ray-tracing procedures and computation of Lyapunov exponents and, in particular, running away for explicit NED sources supporting any such geometries and modifying light propagation in agreement with the comments of our footnote 1.

Our list is thus comprised of the following BH geometries (for further details on these geometries see the references \cite{Vagnozzi:2022moj,daSilva:2023jxa}):
\begin{enumerate}[i)]
    \item \textbf{Reissner-Nordstr\"om (RN) BH}
    \begin{equation}
            A(r) = 1-\frac{2M}{r} + \frac{q_e^2}{r^2} \ .
    \end{equation}
    This is the standard static Maxwell-Einstein solution with an electric charge, $q_e$.

    \item \textbf{Bardeen's BH} \cite{Bardeen}
    \begin{equation}
            A(r) = 1-\frac{2Mr^2}{(r^2+q_m^2)^{3/2}} \ .
    \end{equation}
    This solution is a non-singular BH. It includes an effective regularization parameter, $q_m$, to avoid the central singularity, nowadays understood as a magnetic charge.

    \item \textbf{Hayward BH} \cite{Hayward:2005gi}
    \begin{equation}
            A(r) = 1-\frac{2Mr^2}{r^3+2l^2M} \ .
    \end{equation}
     Similar to Bardeen's solution, the Hayward metric is a non-singular BH geometry arising from various regularization mechanisms, with $l$, its regularization parameter, associated to a length scale.
    \item \textbf{Frolov BH} \cite{Frolov:2016pav}
    \begin{equation}
            A(r) = 1-\frac{(2Mr-q_e^2)r^2}{r^4+(2Mr+q_e^2)l^2} \ .
    \end{equation}
    This model includes both electric charge, $q_e$, and a regularization parameter, $l$, allowing the study of the effect of charge and singularity regularization in BHs.

    \item \textbf{Kazakov-Solodukhin (KS) regular BH} \cite{Kazakov:1993ha}
    \begin{equation}
        A(r)=-\frac{2M}{r} + \frac{\sqrt{r^2-l^2}}{r} \ .
    \end{equation}
    This is another non-singular BH that includes quantum-gravity corrections in the form of a length scale, $l$.

    \item \textbf{Sen BH} \cite{Sen:1992ua}
    \begin{equation}
        A(r)=1-\frac{2M}{r+q_m^2/M} \ .
    \end{equation}
    This BH is motivated by solutions derived from low energy effective field theory in heterotic string theory. It includes an effective charge parameter $q_m$.

    \item \textbf{Einstein-Maxwell-Dilation (EMD) BH} \cite{Gibbons:1987ps}
    \begin{equation}
        A(r)=1-\frac{2M}{r} \left( \sqrt{1+\frac{q_e^4}{4M^2 r^2}} - \frac{q_e^2}{2Mr} \right) \ .
    \end{equation}
    This model results from coupling gravity to a dilaton field, which modifies the standard Einstein-Maxwell action. The parameter, $q_e$, defines the effect of the dilaton field.

    \item \textbf{Dark matter (DM)-surrounded BH} \cite{Li:2012zx}
    \begin{equation}
        A(r)=1-\frac{2M}{r} +\frac{k}{r} \log \left(\frac{r}{\vert k \vert} \right) \ .
    \end{equation}
    This geometry models the effect of a surrounding dark matter halo, described by $k$, the DM coupling constant.

    \item \textbf{Conformal scalar model (ConfSca) BH} \cite{Astorino:2013sfa}
    \begin{equation}
        A(r)=1 - \frac{2M}{r} -\frac{s}{r^2} \ .
    \end{equation}
    This solution arises from a conformal coupling between a scalar field and the gravitational field. The parameter, $s$, is the scalar-gravity coupling constant.

    \item \textbf{Ghosh-Kumar (GK) BH} \cite{Ghosh:2021clx}
    \begin{equation}
        A(r)=1-\frac{r_S}{\sqrt{r^2+q_m^2}} \ .
    \end{equation}
    This solution emerges from coupling GR to a NED. The NLED parameter, $q_m$, describes the coupling.
\end{enumerate}

\subsection{Constraints from EHT observations}\label{S:eht_constraints}

\begin{figure*}[t!]
    \centering
    \includegraphics[width=0.95\textwidth]{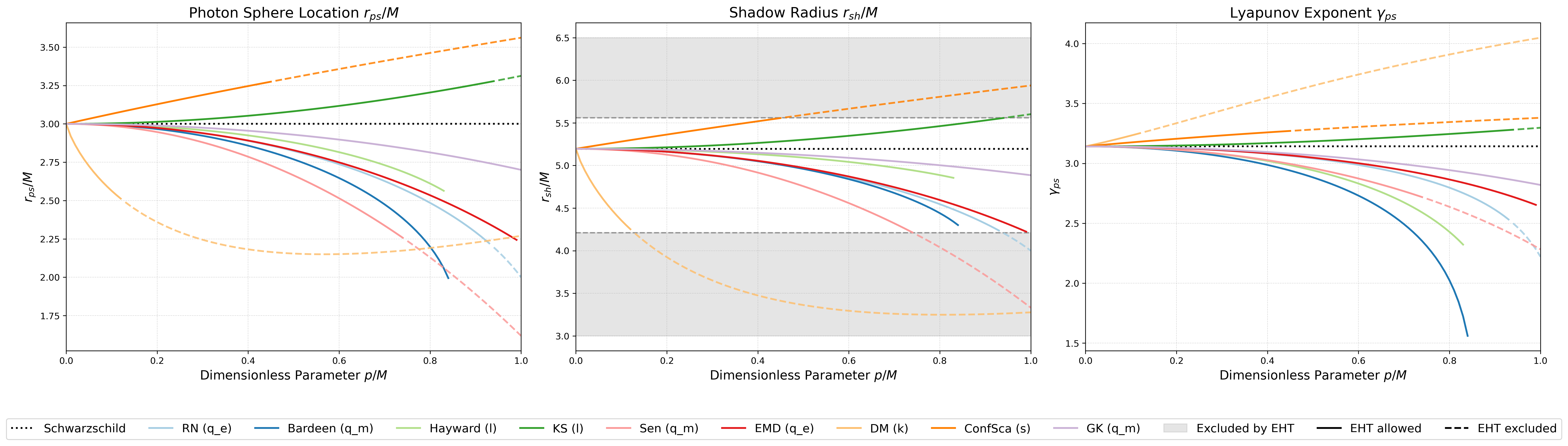}
    \hfill
    \includegraphics[width=0.95\textwidth]{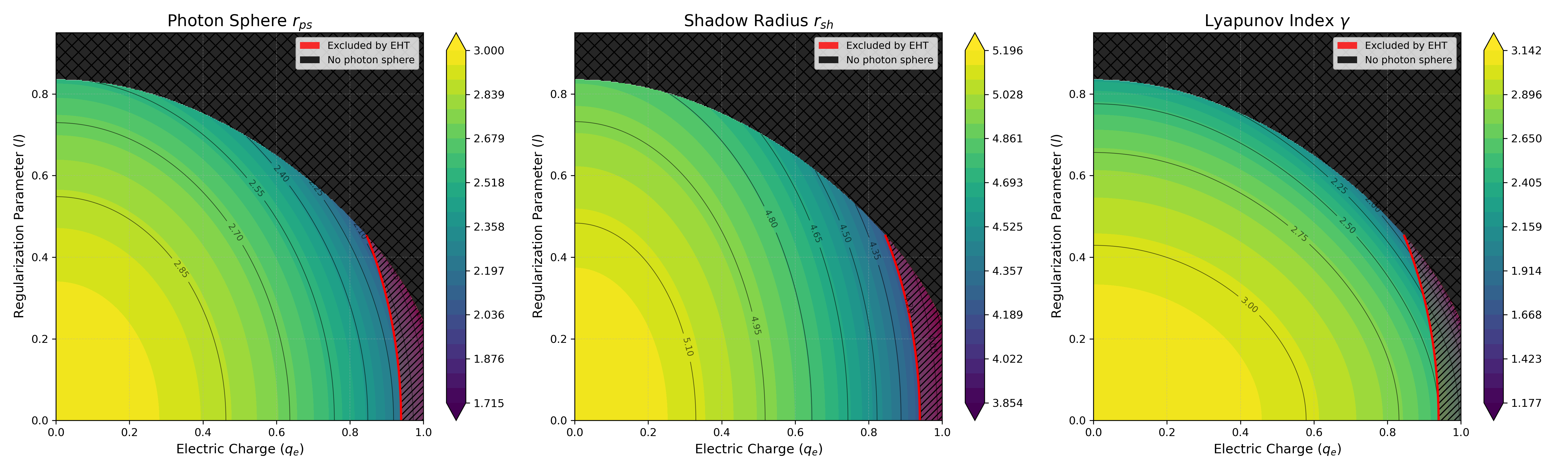}

    \caption{From left to right: the location of the photon sphere, the radius of the shadow (i.e. the critical impact parameter) and the Lyapunov exponent for the parameters of each model.
    (a) On the top plot, we compare the models with only one parameter. We plot as a solid (dashed) line the allowed (excluded) by EHT constraints. We shade the excluded region on the shadow radius plot. 
    (b) On the bottom plot, we plot the Frolov metric as it includes two parameters. We also shade in red the excluded EHT region. We shade in black the region where there is not a photon sphere. All the parameters are scaled with the mass $M=1$. 
    }
    \label{fig:combined_results}
\end{figure*}

Each of the models introduced in the previous subsection is characterized by one or more parameters that quantify its deviation from the Schwarzschild geometry. For most of the models, which depend on a single parameter, we analyze the behavior of the relevant BHI observables by treating this parameter as a generic, dimensionless quantity $p/M$. This allows a direct comparison across different theoretical frameworks, illustrating how the photon sphere radius $r_{ps}$, the critical impact parameter $b_{ps}$ (or shadow radius), and the Lyapunov exponent $\gamma_{ps}$ vary as a function of the parameter that controls its departure from the Schwarzschild limit ($p/M=0$). We consider the range of validity of each model to obtain physical solutions, e.g. the Reissner-N\"ordstr\"om solution yields a naked singularity when $q_e\geq M$.

One of the tools we have at our disposal to constraint the parameters of such modified BHs are the EHT imaging of M87$^*$ and Sgr A$^*$ and, in particular, the inference on the shadow's radius from \cite{EventHorizonTelescope:2022xqj}. It should be pointed out that shadow's size measurements face the difficulty that the EHT Collaboration techniques cannot measure brightness contrasts below $\sim 10\%$ of the peak brightness. Nonetheless, using a methodology that relies on a correlation between the observed angular size of the bright ring of radiation, and the theoretically-computed size of the shadow via $b_c$ (and after proper calibration accounting for theoretical and measurement biases) they delivered constraints on the fractional deviation in the shadow's radius as compared to the Schwarzschild expectations. The net result is that, at $2\sigma$, the shadow's radius is constrained as \cite{Vagnozzi:2022moj}
\begin{equation} \label{eq:bounsha}
    4.21 \leq \frac{r_{sh}}{M} \leq 5.56,
\end{equation}
Recall that in the case of the Schwarzschild geometry, $r_{sh}/M =3\sqrt{3} \approx 5.196$. 

It should be pointed out that the above correlation and, consequently, the constraint~\eqref{eq:bounsha}, is only valid under specific astrophysical assumptions. Nonetheless, for the sake of our analysis we shall take such a constraint at face value and use it as a boundary condition on the allowed parameter space, instead of a fundamental limit. In this sense, the location of the photon sphere, $r_{ps}/M$, the shadow radius, $r_{sh}/M$, and the Lyapunov exponent, $\gamma_{ps}$, for each of the models, are presented (from left to right in the top plots) in Fig.~\ref{fig:combined_results}. For the two-parameter Frolov model, we show (bottom plots) a 2D contour plot with both free parameters (scaled with the mass) as variables. For the single-parameter models, the plot for the shadow radius provides the allowed parameter-space (white region) according to the EHT constraints (\ref{eq:bounsha}).

\subsection{The correspondence in action: computation of QNM frequencies and BHI quantities}

We now investigate the core of the question: does the QNM-BHI correspondence allow for a faithful reproduction of the relevant quantities (QNM real and imaginary frequencies, and critical impact parameter and Lyapunov index, respectively) on one side of it starting from the other?. To investigate this question we proceed as follows. 
For each model, we select specific parameters defined by the saturation of the EHT constraints in Sec.~\ref{S:eht_constraints}, found via the top-middle plot of Fig.~\ref{fig:combined_results}. Then, we compute the spectrum of QNM modes for scalar fields at the fundamental mode $n=0$ and at large multipole number $\ell=10$, finding the QNM frequency's real and imaginary parts. Next we employ the correspondence to find the expected critical impact parameter and Lyapunov index on the BHI side. Then we use the results of \cite{daSilva:2023jxa} regarding the values of such quantities from direct computation using BHI to find the ratio between the critical impact parameters and the Lyapunov indices computed from using the methods of QNM and BHI. Finally, we repeat the exercise as lower values of $\ell$ are considered until arriving to $\ell=2$. The results are displayed in Table \ref{tab:1}, making explicit the values of the parameters employed, and where we also compute the result for the Schwarzschild geometry as the benchmark solution.

As one could expect from the inner workings of the correspondence, for large values of $\ell$, such as $\ell=10$, the correspondence works extremely well, with degrees of agreement both for the ratio between critical impact parameter and Lyapunov indices derived from each messenger (i.e. directly from BHI and indirectly via the correspondence) that are very close to unity. Some BH geometries display tiny modifications attributed to inefficiencies in the numerical procedures to finding the QNM frequencies. What is quite surprising is the fact that for lower values of $\ell$, even when getting to $\ell=2$, the correspondence still works very nicely at every geometry considered. We point out that methods for computation of quantities on each side of the correspondence are not correlated. For BHI we simply use the expression (\ref{eq:lyaexpp}) of the Lyapunov index, evaluated at the photon sphere as provided by the conditions (\ref{eq:cir1}) and (\ref{eq:cir2}). As for QNMs, the WKB method applied to the wave equation (\ref{eq:weq}) is employed. Therefore, this is a genuine effect of the correspondence to cross the divide between each messenger.

We can now elaborate on plausible reasons behind the effectiveness of the correspondence at low $\ell$. We first note that the models considered here satisfy the condition $  g_{tt} g_{rr} = -1  $, significantly simplifying the radial equation for scalar test field perturbations. This generates a mathematically well-behaved effective potential, which includes the appropriate behavior of $  V(r)  $ at the horizon and at infinity. Thus, for the simple cases considered, the WKB matching between the asymptotic regions (plane waves) and the Taylor expansion around the maximum of the potential can be performed perfectly. No spurious errors arise, and the frequencies obtained are the correct ones. In the eikonal limit, this is precisely the regime where the geometric physics of unstable null orbits manifests itself clearly. However, when higher-order corrections are considered in the WKB method, this generally implies a significant improvement in the accuracy of the obtained frequencies, as occurs in the cases we have computed. In this sense, we considered here the 6th order of the WKB approximation because it is known that this order generally produces the best results.

The smoothness of the effective potentials chosen around their single-peak thus implies that even for low $\ell$ the corresponding QNMs only probe a small neighborhood around the potential's maximum and, if the such a neighborhood is well approximated by an inverted parabola, then only its peak and curvature are relevant. Since these are the quantities that enter into the photon sphere and Lyapunov exponent, this might explain why the correspondence still retains validity in such a low $\ell$ regime. On the other hand, here we are considering the fundamental mode, for which the condition $  \ell \gtrsim n  $ is satisfied, representing the regime where the WKB approximation works best. In the case that the effective potential has multiple peaks or long tails the correspondence should deteriorate much faster. This would imply that the usefulness of the correspondence at low $\ell$ is confined to a restricted class of (spherically symmetric), smooth-enough geometries.

\begin{table*}[h!]
    \centering
    \renewcommand{\arraystretch}{1.1}
    \resizebox{1.6\columnwidth}{!}{
        \begin{tabular}{|c|c|c|c|c|c|c|c|c|c|}
        \hline

        \textbf{Model} & \textbf{$A(r)$} & \textbf{$b_{\mathrm{ps}}^{\mathrm{BHI}}$} & \textbf{$\gamma_{\mathrm{ps}}^{\mathrm{BHI}}$} & \textbf{$\ell$} & \textbf{\text{QNMs} ($\omega_R +i \omega_I$)} & \textbf{$b_{\mathrm{ps}}^{\mathrm{QNM}}$} & \textcolor{blue}{$\Delta b_{\mathrm{ps}} (\%)$} & $\gamma_{\mathrm{ps}}^{\text{QNM}}$ & \textcolor{blue}{$\Delta\gamma_{\mathrm{ps}} (\%)$} \\
        \hline

        \hline
        & & & & 10 & $2.021320 -0.096256\, i$ & 5.195 & 0.019 & 3.142 & 0.000 \\
        & & & & 8 & $1.636560 -0.096272\, i$ & 5.194 & 0.038 & 3.142 & 0.000 \\
        Schwarzschild & $1-\frac{2M}{r}$ & 5.196 & 3.142 & 6 & $1.251890 -0.096310\, i$ & 5.192 & 0.077 & 3.142 & 0.000 \\
        & & & & 4 & $0.867400 -0.096390\, i$ & 5.188 & 0.154 & 3.142 & 0.000 \\
        & & & & 2 & $0.484000 -0.097000\, i$ & 5.165 & 0.597 & 3.148 & 0.191 \\
        \hline

        \hline
        & & & & 10 & $2.494810 -0.094645\, i$ & 4.209 & 0.000 & 2.503 & 0.950 \\
        & & & & 8 & $2.019909 -0.094653\, i$ & 4.208 & 0.024 & 2.503 & 0.950 \\
        RN & $1-\frac{2M}{r} + \frac{q_e^2}{r^2}$ & 4.209 & 2.527 & 6 & $1.545111 -0.094670\, i$ & 4.207 & 0.048 & 2.502 & 0.990 \\
        $q_e = 0.939$ & & & & 4 & $1.070553 -0.094714\, i$ & 4.203 & 0.143 & 2.501 & 1.029 \\
        & & & & 2 & $0.596822 -0.094905\, i$ & 4.189 & 0.475 & 2.498 & 1.148 \\
        \hline

        \hline
        & & & & 10 & $2.321584 -0.077895\, i$ & 4.523 & 0.022 & 2.214 & 1.731 \\
        & & & & 8 & $1.879547 -0.077908\, i$ & 4.522 & 0.044& 2.214 & 1.731 \\
        Bardeen & $1-\frac{2Mr^2}{(r^2+q_m^2)^{3/2}}$ & 4.524 & 2.253 & 6 & $1.437568 -0.077936\, i$ & 4.522 & 0.044 & 2.214 & 1.731 \\
        $q_m = 0.7698$ & & & & 4 & $0.995722 -0.078008\, i$ & 4.519 & 0.111 & 2.214 & 1.731 \\
        & & & & 2 & $0.554324 -0.078332\, i$ & 4.510 & 0.309 & 2.220 & 1.465 \\
        \hline

        \hline
        & & & & 10 & $2.135769 -0.081463\, i$ & 4.916 & 0.000 & 2.516 & 1.023 \\
        & & & & 8 & $1.729082 -0.081466\, i$ & 4.916 & 0.000 & 2.516 & 1.023 \\
        Hayward & $1-\frac{2Mr^2}{r^3+2l^2M}$ & 4.916 & 2.542 & 6 & $1.322440 -0.081474\, i$ & 4.913 & 0.061 & 2.516 & 1.023 \\
        $l = \sqrt{16/27}$ & & & & 4 & $0.915908 -0.081501\, i$ & 4.916 & 0.000& 2.516 & 1.023 \\
        & & & & 2 & $0.509747 -0.081723\, i$ & 4.904 & 0.244 & 2.518 & 0.944 \\
        \hline

        \hline
        & & & & 10 & $2.451426 -0.088177\, i$ & 4.283 & 0.000 & 2.373 & 1.249 \\
        & & & & 8 & $1.984639 -0.088184\, i$ & 4.283 & 0.000 & 2.373 & 1.249 \\
        Frolov & $1-\frac{(2Mr-q_e^2)r^2}{r^4+(2Mr+q_e^2)l^2}$ & 4.283 & 2.403 & 6 & $1.517906 -0.088198\, i$ & 4.282 & 0.023 & 2.373 & 1.249 \\
        $q_e = 0.875$ & $l = 0.3$ & & & 4 & $1.051298 -0.088236\, i$ & 4.280 & 0.070 & 2.373 & 1.249 \\
        & & & & 2 & $0.585116 -0.088429\, i$ & 4.273 & 0.233 & 2.374 & 1.207 \\
        \hline

        \hline
        & & & & 10 & $1.889278 -0.093975\, i$ & 5.558 & 0.018 & 3.282 & 0.182 \\
        & & & & 8 & $1.529635 -0.093993\, i$ & 5.557 & 0.036 & 3.282 & 0.182 \\
        KS & $-\frac{2M}{r} + \frac{\sqrt{r^2-l^2}}{r}$ & 5.559 & 3.288 & 6 & $1.170067 -0.094029\, i$ & 5.555 & 0.072 & 3.282 & 0.182 \\
        $l = 0.942$ & & & & 4 & $0.810674 -0.094123\, i$ & 5.551 & 0.144 & 3.282 & 0.182 \\
        & & & & 2 & $0.451882 -0.094527\, i$ & 5.532 & 0.486 & 3.286 & 0.061 \\
        \hline

        \hline
        & & & & 10 & $2.303943 -0.100348\, i$ & 4.557 & 0.022 & 2.873 & 0.485 \\
        & & & & 8 & $1.865402 -0.100362\, i$ & 4.557 & 0.022 & 2.873 & 0.485 \\
        Sen & $1-\frac{2M}{r+q_m^2/M}$ & 4.558 & 2.887 & 6 & $1.426966 -0.100391\, i$ & 4.555 & 0.066 & 2.873 & 0.485 \\
        $q_m = 0.6$ & & & & 4 & $0.988774 -0.100467\, i$ & 4.551 & 0.154 & 2.873 & 0.485 \\
        & & & & 2 & $0.551416 -0.100789\, i$ & 4.534 & 0.527 & 2.871 & 0.554 \\
        \hline

        \hline
        & & & & 10 & $2.494056 -0.100040\, i$ & 4.210 & 0.024 & 2.646 & 0.713 \\
        & & & & 8 & $2.019331 -0.100053\, i$ & 4.210 & 0.024 & 2.646 & 0.713 \\
        EMD & $1-\frac{2M}{r} \left( \sqrt{1+\frac{q_e^4}{4M^2 r^2}} - \frac{q_e^2}{2Mr} \right)$ & 4.211 & 2.665 & 6 & $1.544719 -0.100078\, i$ & 4.208 & 0.071 & 2.646 & 0.713 \\
        $q_e = 0.995$ & & & & 4 & $1.070372 -0.100143\, i$ & 4.204 & 0.166 & 2.645 & 0.750 \\
        & & & & 2 & $0.596926 -0.100421\, i$ & 4.188 & 0.546 & 2.643 & 0.826 \\
        \hline

        \hline
        & & & & 10 & $2.493248 -0.123217\, i$ & 4.211 & 0.071 & 3.260 & 0.245 \\
        & & & & 8 & $2.018681 -0.123240\, i$ & 4.209 & 0.000 & 3.259 & 0.275 \\
        DM & $1-\frac{2M}{r} +\frac{k}{r} \log \left(\frac{r}{\vert k \vert} \right)$ & 4.212 & 3.268 & 6 & $1.544228 -0.123287\, i$ & 4.209 & 0.071 & 3.260 & 0.245 \\
        $k = 0.128$ & & & & 4 & $1.070046 -0.123412\, i$ & 4.205 & 0.166 & 3.261 & 0.214 \\
        & & & & 2 & $0.596790 -0.123950\, i$ & 4.189 & 0.546 & 3.262 & 0.184 \\
        \hline

        \hline
        & & & & 10 & $1.890103 -0.093712\, i$ & 5.555 & 0.018 & 3.271 & 0.214 \\
        & & & & 8 & $1.530305 -0.093729\, i$ & 5.554 & 0.036 & 3.271 & 0.214 \\
        ConfSca & $1 - \frac{2M}{r} -\frac{s}{r^2}$ & 5.556 & 3.278 & 6 & $1.170582 -0.093764\, i$ & 5.553 & 0.054 & 3.271 & 0.214 \\
        $s = 0.45$ & & & & 4 & $0.811037 -0.093856\, i$ & 5.548 & 0.144 & 3.272 & 0.183 \\
        & & & & 2 & $0.452102 -0.094250\, i$ & 5.530 & 0.468 & 3.275 & 0.092 \\
        \hline

        \hline
        & & & & 10 & $2.491058 -0.077583\, i$ & 4.215 & 0.024 & 2.055 & 2.143\\
        & & & & 8 & $2.016925 -0.077602\, i$ & 4.214 & 0.047 & 2.055 & 2.143 \\
        GK & $1-\frac{2M}{\sqrt{r^2+q_m^2}}$ & 4.216 & 2.100 & 6 & $1.542911 -0.077640\, i$ & 4.213 & 0.071 & 2.055 & 2.143 \\
        $q_m = 1.63$ & & & & 4 & $0.596308 -0.078081\, i$ & 4.192 & 0.569 & 2.057 & 2.048 \\
        & & & & 2 & $0.361459 -0.078447\, i$ & 4.150 & 1.565 & 2.045 & 2.619 \\
        \hline

        \end{tabular}
    }
    \caption{Comparison of the values of the critical impact parameter $b_{\mathrm{ps}}^{\mathrm{BHI}}$ and the Lyapunov exponent $\gamma_{\mathrm{ps}}^{\mathrm{BHI}}$ found from direct BHI procedures, and from the analysis of massless scalar QNMs, $b_{\mathrm{ps}}^{\mathrm{QNM}}$ and $\gamma_{\mathrm{ps}}^{\mathrm{QNM}}$,
    \textcolor{blue}{
    using the sixth-order WKB approximation
    }
    for the fundamental mode $n=0$. The pool of spherically symmetric BH geometries is a restriction of those considered in \cite{daSilva:2023jxa}, and their parameter(s) are constrained according to the EHT observations on the shadow's size of M87$^*$ and Sgr A$^*$ \cite{EventHorizonTelescope:2022xqj} such that each model saturates the corresponding bound. Our analysis sweeps over several values of the angular number $\ell$ (assuming $\ell > n$). The latter analysis provides the real and imaginary parts of the frequency, which are inserted into the correspondence relations (\ref{eq:correspondence1}) and (\ref{eq:correspondence2}) and using Eq.(\ref{eq:lyaprin}) to provide their predictions for the corresponding $b_{\mathrm{ps}}^{\mathrm{QNM}}$ and the Lyapunov exponent $\gamma_{\mathrm{ps}}^{\mathrm{QNM}}$. \textcolor{blue}{In this Table $\Delta b_{\text{ps}} \equiv 100 \vert b_{\text{ps}}^{\text{BHI}}-b_{\text{ps}}^{\text{QNM}} \vert/b_{\text{ps}}^{\text{BHI}}$ and $\Delta \gamma_{\text{ps}} \equiv 100 \vert \gamma_{\text{ps}}^{\text{BHI}}-\gamma_{\text{ps}}^{\text{QNM}} \vert/\gamma_{\text{ps}}^{\text{BHI}}$ are the percentage relative error between quantities computed on each side of the correspondence.}}
    \label{tab:1}
\end{table*}

\subsection{Usefulness of the correspondence vs caveats on QNM and BHI sides}

In the multimessenger era, the fact that the QNM-BHI correspondence enables a connection between gravitational-wave observations (LIGO/Virgo/KAGRA, LISA, Einstein Telescope) and electromagnetic imaging (EHT, ngEHT, Black Hole Explorer) could provide complementary insights into near-horizon physics \cite{Cabero:2019zyt}. Consequently, any modification to the QNM spectrum induced by putative deviations in the underlying space-time geometry should be correlated with modifications to quantities of the BHI, and vice versa. In this regard, recent works have exploited this connection to verify analytically and numerically the QNM-BHI correspondence in GR and modified gravity theories \cite{Guo:2021enm,Konoplya:2022gjp,Konoplya:2022tvv,Hashimoto:2023buz,Giataganas:2024hil,Chen:2022nlw}. However, such expectations are subjected to several caveats regarding the connection between theoretical quantities and actual observables for each messenger.

On the QNM side, the correspondence puts forward the relevance of the real and imaginary parts of their frequency which are, in general, difficult to be calculated by the existing methods, all of them showing some kind of weakness. For instance, the WKB method, which is one of the most widely used in the literature (and the one employed here) only works reliably for smooth, single-peaked potentials, struggling to deal with more complicated geometries \cite{Iyer:1986np}, while more precise methods such as continued-fraction or spectral ones are mathematically and numerically more complex while suffering from convergence and stability issues and being less flexible to deal with complicated space-times \cite{Berti:2009kk}. This way, the correspondence provides a new way of calculating the QNM frequencies which can be later be double-checked with other methods, providing stronger and more reliable results.  

Observationally, the real and imaginary parts of QNM frequencies are inferred from measurements of the GW signal following a BH merger. Within this framework, BH spectroscopy treats the signal as a superposition of damped oscillatory models, where the mode frequencies and damping times are governed by the QNM spectrum of the resulting BH, while each mode's amplitude and phases are determined from the non-linear merger dynamics. Importantly, these physical quantities are not measured directly but instead extracted from the data via different methods, such as fitting or Bayesian parameter estimation \cite{Carullo:2019flw}.  In practice, this is a challenging task due to overlapping of different modes, detector noise, low signal-to-noise ratio, and ambiguities in defining where the ringdown phase begins \cite{LIGOScientific:2025wao}. Indeed, the event GW250114 that has been the loudest to date is the only one capable of extracting the first overtone of the quasi-normal spectrum at a certain reasonable confidence level of $4.1\sigma$, and which is consistent with the Kerr assumption on the nature of the final object \cite{LIGOScientific:2025rid}.

On the BHI side, the correspondence gives relevance to the critical impact parameter and the Lyapunov exponent, respectively, but these are not direct observables. The critical curve delineates a region of highly-bent trajectories that are blend into a single photon ring whose inner edge closely tracks such a curve, such that its inner region presents a marked brightness deficit due to the much shorter path-length of those photons intersecting the event horizon. The shadow defined this way via the critical impact parameter, however, only occurs in the case of spherical accretion \cite{Bauer:2021atk}. Indeed, for geometrical configurations of a disk with gaps in the emitting region and displaying some optical thinness, highly-bent light trajectories are decomposed into an infinite sequence of $n$ photon rings nested in the disk's direct emission and  also approaching, in the limit $n \to \infty$, the critical curve \cite{Chael:2021rjo,Vincent:2022fwj,Cardenas-Avendano:2023dzo}. Every single photon ring now appears as local brightness enhancements. Furthermore, if the inner edge of the disk extends to regions near the BH event horizon, the central brightness depression is strongly reduced as compared to the canonical shadow, breaking this correlation between dark central region, shadow's radius, and (via the correspondence) the QNM amplitude. Without detailed information on the disk properties surrounding an imaged BH, one cannot observationally determine where the critical curve is in a reliable way. 

On the other hand, testing the Lyapunov index requires the measurement of two successive rings and application of the following relations satisfied by them
\cite{Johnson:2019ljv,Kocherlakota:2023qgo}
\begin{equation} \label{eq:scaleprs}
\frac{b_{n+1} - b_{ps}}{b_n -  b_{ps}} \approx \frac{\omega_{n+1}}{\omega_n} \approx \frac{I_{\nu;n+1}}{I_{\nu;n}} \approx e^{-\gamma_{ps}} 
\end{equation}
where $\{\eta_n,\omega_n,I_{\nu;n}\}$ are the $n^{th}$ image radii, width, and flux intensity, respectively. While such relations only make full sense in the limit $n \to \infty$, they provide very good approximations even at low $n$'s, which allow us to approximate the Lyapunov index via the ratio between the $n=2$ and $n=1$ rings. However, such a measurement is subject to several caveats. First of all, the Lyapunov index does not account for the local differences in the emissivity of the plasma as seen by the traveling photons belonging to each photon ring, meaning that modeling of the emitting plasma will introduce deviations in the actual ratios between photon rings as compared to Lyapunov-based expectations. Observationally, the enhancement of brightness produced by photon rings leaves a characteristic signal when transformed to the Fourier space, which is what very long baseline interferometric projects (like the EHT)  measure, and whose periodicity can be used to provide a diameter of the corresponding ring \cite{Gralla:2020nwp}. Future interferometric prospects, like the ngEHT or the Black Hole Explorer, are expected to measure the $n=1$ photon ring. While an accurate measurement of the $n=2$ ring seems unfeasible in the near feature, the community is exploring further tools to its analysis such as for auto-correlations in the brightness fluctuations of the photon rings' emission \cite{Hadar:2020fda}, tracking of dynamical, flaring events or hot-spots  \cite{Broderick:2005my,Kocherlakota:2024hyq}.

\section{Conclusion and discussion} \label{S:VI}

The emergence of multimessenger astronomy, which uses different carriers to probe the universe, offers the exciting opportunity to study astronomical objects through the combined analysis of multiple such messengers. Regarding BHs, we currently have at our disposal both GWs in the ringdown phase of coalescence events of binary systems, and optical imaging of the accretion flow surrounding supermassive BHs. Each messenger therefore explores different ranges of masses and curvatures and equally importantly, different aspects (dynamical vs quasi-static) of the BH geometry. The information provided by QNM and BHI can be employed either as null-tests of the Kerr hypothesis, or to seek traces of non-Kerrness, i.e., modified BHs different from Kerr or even horizonless compact objects. 

A growing trend in the literature is to find useful tools allowing the information collected from each messenger to be combined with each another in order to further constrain the nature of the BH. This is the case of the QNM-BHI correspondence put forward in the recent literature, and which relates the real and imaginary parts of the QNM frequency spectrum with the critical impact parameter and Lyapunov index of nearly-bound geodesics in the BHI side. This way, the correspondence bridges GW spectroscopy and electromagnetic observations, offering us new channels for geometric probes of near-horizon physics.

In this work we have discussed the accuracy and caveats of the QNM-BHI correspondence using several spherically symmetric geometries describing modified BHs that have attracted certain attention in the literature in the last few years. We first used previous constraints on their space of parameters based on the inference made by the EHT Collaboration on their shadow's size. Subsequently we have made use of the WKB method to compute the QNM frequencies for $n=0$ and different values of the multipole number $\ell$ and from there, using the correspondence, verify their accuracy in predicting the right quantities on the BHI side. Our results put forward that not only is the correspondence precise at large values of $\ell$ (say $\ell =10$), as expected, but also retains large accuracy for low values of $\ell$. The reasons behind why the correspondence is still useful at low $\ell$ seem to be related to the single-peaked, smooth character of the effective potential for the spherically symmetric configurations considered in this work, requiring a more systematic analysis of the conditions under which it will continue to be so for a given space-time metric.

While the above result seems to give a strong usefulness to the correspondence for certain classes of space-time configurations, we have also made a critical assessment on the caveats regarding the connection between theoretical quantities and observational signatures on each side of the correspondence. One immediate application of our analysis is that, given the fact that BHI quantities are far more simpler to be obtained (since they only involve algebraic manipulations and equation-solving) than the QNM ones (which requires the application of numerical recipes whose accuracy and reliability is not always easy to guarantee), then BHI may prove more useful to safeguard computation of QNMs. On the other hand, the addition of the analysis of greybody factors in scattering problems, which can also be framed within the correspondence \cite{Pedrotti:2025idg}, could provide further useful tools for the sake of each messenger.

At present, we are far from being able to test the same BH using both messengers, since the conditions under which a BH can be probed by one messenger are not typically favorable to the other \cite{Yang:2021zqy}. The closest setup we have currently within our reach is that of extreme mass-ratio inspirals, which are expected to be measured with LISA. The fact that these systems should be accompanied by an accretion disks offers a potential opportunity to run tests with BHI \cite{Pan:2021ksp}, though difficulties have been pointed out regarding the fact that their faint signal make them poorly suitable to carry out GW spectroscopy \cite{Baibhav:2019rsa}.  Whether the correspondence might prove to be useful within this context requires still more work regarding its inner workings and potential contributions to the phenomenology.

\section*{Acknowledgments}

A.~R. would like to express his gratitude to Silesian University in Opava, Czech Republic for their financial support, and is very grateful for the hospitality of the University of Valencia (Spain), Valencia Polytechnic University (Spain) and
the Complutense University of Madrid (Spain). This work is supported by the Spanish National Grants PID2022-138607NB-I00, PID2024-157196NB-I00, and CNS2024-154444, funded by  MICIU/AEI/10.13039/501100011033; the Department of Education, Junta de Castilla y Le\'on and FEDER Funds, Ref. CLU-2023-1-05; and the grant program Vouchers for Universities in the Moravian-Silesian Region (registration number CZ.10.03.01/00/23\_042/00003901119). This article is based upon work from COST Action FuSe, CA24101, supported by COST (European Cooperation in Science and Technology).

\end{document}